\documentclass[usenatbib]{mn2e}
\usepackage{amsmath}
\usepackage{graphicx}
\newcommand{\mc}[1]{{\mathcal{#1}}}

\newcommand{\be}{\begin{equation}}
\newcommand{\ee}{\end{equation}}

\begin{document}
\title[The Virtues of Frugality]{The Virtues of Frugality -- Why cosmological  observers should release their data slowly}
\date{\today} 

\author[G.D.~Starkman et al]{Glenn D.~Starkman$^{1}$, Roberto Trotta$^{2}$ and Pascal M.~Vaudrevange$^{1}$\\
$^{1}$CERCA \& Department of Physics, Case Western Reserve University, 10900 Euclid Ave, Cleveland, OH 44106, USA\\
$^{2}$Astrophysics Group, Imperial College London, Blackett Laboratory, Prince Consort Road, London SW7 2AZ, UK}

\maketitle

\begin{abstract}
Cosmologists will soon be in a unique position. Observational noise
will gradually be replaced by cosmic variance as the dominant source
of uncertainty in an increasing number of observations.  We reflect on the
ramifications for the discovery and verification of new models.  If
there are features in the full data set that call for a new model,
there will be no subsequent observations to test that model's
predictions.  We give specific examples of the problem by discussing the pitfalls of model discovery by prior adjustment in the context of dark energy models and inflationary theories. We show how the gradual release of data can mitigate
this difficulty, allowing anomalies to be identified, and new models
to be proposed and tested.  We advocate that observers plan for the
frugal release of data from future cosmic variance limited
observations.
\end{abstract}

\section{Introduction} 

Cosmologists can only make observations on 
(or occasionally within) our
past light cone. Whatever the reality of the multiverse, we
Earth-bound humans of the $T_{CMB}\!=\!2.725K$ era  have access
to only a finite volume of space, containing finite energy and
information.  The exciting period in which we find ourselves learning
more and more about this volume of accessible space and its contents
cannot last forever.  While we are unlikely to gather {\em all} the
existing information content of the observable universe, we are
already making substantial inroads on the information of cosmic
significance.

The most notable example of confronting the finite information content
of the universe is our measurements of the power in the lowest
multipoles $C_{\ell}$ of the cosmic microwave background (CMB)
temperature anisotropies.  Their statistical error bars are now
smaller than the ``cosmic variance" errors -- the expected difference
between what we measure for these multipoles and what we would measure
if we could average over many independent horizon volumes. The range
of $\ell$ for which this is true is increasing as the Wilkinson
Microwave Anisotropy Probe (WMAP) continues to report new results.
This trend will accelerate as new experiments join the fray. (Though
we could wait a few hundred million years to gain access to a
mostly-independent last scattering surface.)

The CMB temperature-temperature power spectrum is unlikely to be the
last place where the finite universe limits cosmology.  Astronomical
surveys are already cataloguing an increasing fraction of all the
structures within our past light cone.  Redshifted hydrogen hyperfine
instruments will eventually extend the volume over which we map the
structure of matter nearly out to the horizon.

There are consequences to becoming a data limited science. We upset
the balance between applying the brain's remarkable pattern-finding
abilities and testing the robustness of the patterns we discover. We
may see patterns in finite data, but, unable to collect new data, we
have no way to confirm their reality, missing out on potentially
significant discoveries. We risk falling for what particle
physicists call ``the look elsewhere effect'', i.e.~the spurious
``discovery'' of statistically significant anomalies which are merely
the consequence of performing a large number of tests on the same
data. A small fraction of those are bound to report significant
``evidence'' for unexpected features due to random noise.  Unlike
experimental scientists, we may no longer be able to collect data,
form a new hypothesis, and test its predictions. Our ability to
distinguish between statistical fluctuations and real effects becomes
limited.

Given that the challenge of finite data is upon us, our best hope is
to devise strategies to minimize its effects.  The approach that we
shall explore and advocate is to simulate the cycle of data
acquisition and analysis by being frugal.  By allowing colleagues to
see only subsets of the data, construct hypotheses based on them, then
test those hypotheses on larger subsets, we can aim to avoid 
unexplained anomalies with untestable explanations.

The benefits of frugality arise not from some magical improvements in
the statistical power of the data, but from acknowledging and
mitigating a basic human failing: over-confidence.  Specifically, by
assigning all probability to the set of physical models that we have
thought about and consequently zero probability to all other models,
we ignore that we may not have considered the correct model. Frugality
allows us to redress those wrongs by admitting such models and testing
their predictions on our remaining data.  We examine the effects of
(and several strategies for) dividing cosmological data into several
pieces so that new models can be consistently explored.

\section{Model discovery} 

\subsection{Bayesian model selection and prior updating}

We take a Bayesian outlook on
hypothesis testing, as we believe (and show below) that this closely
reflects the way we think about models. Another reason for being wary
of the usual (frequentist) practice of reporting $p$-values is that
the latter are {\em not} probabilities for hypotheses, despite being
commonly misinterpreted as such~\citep{Sellke:2001,Gordon:2007xm}.  Suppose we have a
model $M_0$ with parameters ${\bf \theta}_0$, that we wish to evaluate
in light of data $d$. Our updated state of belief in the model's
parameters is given by the posterior probability distribution function
(pdf) on ${\bf \theta}_0$, obtained via Bayes' theorem:
\be
p({\bf \theta}_0|d,M_0)=p(d|{\bf
  \theta}_0,M_0) \frac{p({\bf \theta}_0|M_0)}{p(d|M_0)},
 \ee 
   where
$p(d|{\bf \theta_0}, M_0)$ is the likelihood, $p({\bf \theta_0}|M_0)$
the prior on the parameters ${\bf \theta_0}$, and $p(d|M_0)$ is the
marginal likelihood for $M_0$.  Now suppose we notice a feature in the
data that is not reproduced by model $M_0$ (for example by computing
the doubt, as in \cite{Starkman:2008py}). We invent a model $M_1$ with
parameters ${\bf \theta}_1$ as an explanation for said feature and
compute the evidence for both models ($i=0,1$)
\begin{eqnarray}
  p(d|M_i)&=&\int\!d{\bf \theta}_i p(d|{\bf \theta}_i, M_i) p({\bf \theta}_i|M_i)\, .
\end{eqnarray}
Each model's posterior probability in light of $d$ is given by
$p(M_i|d)={p(d|M_i) p(M_i)}\slash{p(d)}$. The {\em ratio} of our
degrees of belief in the models, the Bayes factor
$B_{10}={p(d|M_1)}/{p(d|M_0)}$, penalizes models that are
unnecessarily complex, for example because of an excessive number of
free parameters, automatically encapsulating
Occam's razor~(see e.g.~\cite{Trotta:2005ar,Trotta:2008qt}). In order to increase confidence in the new model $M_1$,
all that is required is $B_{10} > 1$, i.e. that $M_1$ be a more
``effective'' description of the {\it presently} available data. There
is no dependence on the model's predictivity for {\it future}
observations.

In practice, a new model probably would not (and arguably should not)
be accepted until it produces a correct prediction for future data
$d^\prime$ that differs from the old model's, thus enabling the models
to be distinguished.  Formally, the models' relative posterior odds
after seeing both sets of data are given by
\begin{eqnarray}
 \frac{p(M_1|d,d^\prime)}{p(M_0|d,d^\prime)} & = &\frac{p(d^\prime|M_1)}{p(d^\prime|M_0)}\frac{p(d|M_1)}{p(d|M_0)}\frac{p(M_1)}{p(M_0)}\, .
\end{eqnarray}
Before the data set $d$ came along, model $M_1$ was not even on the
table: $p(M_1) = 0$. The step of introducing $M_1$ {\em while
  absolutely crucial}, formally requires the injection of an infinite
amount of information to raise $p(M_1)$ from 0 to a finite value. This
prior adjustment is on top of the change in degree of belief coming
from $d$. It amounts to using the data $d$ twice, first to introduce
$M_1$ by adjusting its prior and then to evaluate the evidence from
$d$.

The duplicate use of the data $d$ leads to posterior odds which can
seriously overstate the statistical significance of a new effect. We
suggest to ``forget'' about the details of $d$, compress its
information into a new non-zero (and still subjective) prior $p(M_1)$,
and then compute the posterior odds arising solely from $d^\prime$,
i.e.
\begin{eqnarray}
 \frac{p(M_1|d, d^\prime)}{p(M_0|d, d^\prime)} & = &\frac{p(d^\prime|M_1)}{p(d^\prime|M_0)}\frac{p(M_1)}{p(M_0)} .
\end{eqnarray}
If an unlimited amount of data is accessible and the anomaly is
correctly modelled by $M_1$, it is {\it guaranteed} to become
eventually favored by the Bayes factor, independent of the exact
choices of priors. Using a finite, cosmic-variance-limited data set
only {\it increases the likelihood} that $M_1$ is confirmed before the
data is exhausted, the more the bigger the fraction of unused data
in $d^\prime$.

\subsection{Examples of prior adjustments in cosmology} 

Two notable
examples in cosmology of devising new models and then adjusting their priors are the
discovery of dark energy and the realization that inflation can easily
accommodate $\Omega<1$.

The discovery of a non-zero, yet tiny cosmological constant $\Lambda$
was in stark contradiction to prior expectations.  Particle-physics
considerations suggested that $\Lambda$ should either be $0$ (model
$\mc{M}_1$) or have a uniform prior between $\pm M_p^4$ (model
$\mc{M}_2$), $p(\Lambda|\mc{M}_1)=\delta(\Lambda),
p(\Lambda|\mc{M}_2)=\Theta(|\Lambda|-M_p^4)\slash{2M_p^4}$, where
$M_p$ is the reduced Planck mass, $\Theta(x)$ is a step function and
$\delta(x)$ is a Dirac delta distribution. Oversimplifying history,
let us assume these were the only theories at hand, and had equal
priors\footnote{An interesting suggestion for choosing model's priors
  based on a Maximum Entropy argument has been put forward
  by~\cite{Brendon}.}: $p(\mc{M}_1)=p(\mc{M}_2)=\frac{1}{2}$.

Along came supernova (SN) redshift measurements
\citep{Perlmutter:1998np}, suggesting a late time acceleration of the
universe driven by (in the simplest models) a small
$\frac{\Lambda_0}{M_p^4}\approx 10^{-120}$. To simplify, let us
assume that the available SN data presented a $5\sigma$ deviation from
$\Lambda=0$.  Computing the Bayes factor using the Savage-Dickey
density ratio \citep{Trotta:2007hy} gives
\begin{eqnarray}
  B_{12}=\frac{p(\Lambda = 0 | d, \mc{M}_2)}{p(\Lambda = 0| \mc{M}_2)}= \frac{10^{121}}{\sqrt{2\pi}} e^{-25/2} \approx10^{115}\, .
  \end{eqnarray}
Due to the strong Occam's razor effect of the prior on $\mc{M}_2$, a
vanishing cosmological constant should hafve still been vastly
preferred, with odds of order $10^{115} : 1$, over a model including a hugely fine-tuned $\Lambda$. A $\sim 23\sigma$ detection of a non-zero cosmological
constant would have been required to override the Occam's razor of the
prior.

However, the particle physics community started reconsidering priors
and developed a new model $\mc{M}_3$ involving anthropic reasoning
which gave more weight to small values of $\Lambda$,
$p(\Lambda|\mc{M}_3)=\Theta(10 \Lambda_0 -
\Lambda)\slash{10\Lambda_0}$, with model priors now
$p(\mc{M}_1)=p(\mc{M}_2)=p(\mc{M}_3)=\frac{1}{3}$. Under the new
anthropic prior, the effect of Occam's razor is vastly reduced, giving
a Bayes factor $B_{13} \approx 10^{-4}$, now favoring model
$\mc{M}_3$. The parameter value that was {\it a priori} considered
unnatural under the original model for a cosmological constant (small
non-zero $\Lambda$) described the data better than the prevailing
model of $\Lambda=0$, but not sufficiently well to be preferred.
Introducing an anthropic model based on the landscape picture in
string theory
\citep{Bousso:2000xa,Giddings:2001yu,Douglas:2003um,Susskind:2003kw,Starkman:2006at}
allowed a small, non-zero cosmological constant to become the
preferred description of the data which has since been supported by
other observations such as CMB and baryon acoustic oscillations.

It is interesting that {\it ex post facto} one might argue that perhaps $\Lambda$ is restricted to be a positive quantity, in which case the appropriate prior would be uniform in $\ln\Lambda$ rather than in $\Lambda$~\citep{Evrard:1995ha,Kirchner}. Under this model $\mc{M}_4$, and assuming a cut-off $\Lambda>\Lambda_{\text{min}}=10^{-500}M_{p}^4$
(see~\cite{Starkman:2006at}), one obtains a Bayes factor $B_{14}\approx 10^{-2}$, i.e.
moderate support for $\Lambda$, analogously to what can be obtained by
anthropic arguments.

An earlier example of discovering a new model through adjusting priors
happened in the mid to late $90$s. The overwhelming evidence for
$\Omega_{\mathrm{tot}} \approx 0.3<1$ posed a problem for inflation,
as it had been viewed to generically predict a flat universe with
$\Omega\approx1$ to high accuracy -- this generally accepted model
could not describe observations. Different models (mostly using
multiple stages of inflation) were devised that produced open
universes \citep{Bucher:1994gb}. In other words, after observing that
$\Omega\approx 0.3$, the priors for single stage inflation, $p(M_0)$,
and for multi stage inflation, $p(M_1)$, were adjusted from $p(M_0)\gg
p(M_1)$ to $p(M_1)\approx p(M_0)$. The prediction for future
observations -- corroborating evidence for $\Omega\approx0.3$ -- was
proven wrong by measurements of $\Omega\approx1$
\citep{Netterfield:2001yq}. The priors were reverted back to $p(M_0)
\gg p(M_1)$, making multi-stage models all but obsolete.

Note that in both the above examples, it was crucial that predictions
of the new model could be tested by follow-up {\em independent}
observations which either confirmed or rejected the new model.

\section{The need for frugality} With the launch of the Planck
satellite, the power spectrum of the temperature fluctuations,
$C_{\ell}^{TT}$, will be limited by cosmic variance all the way up to
$\ell > 2000$. No future observation will ever obtain more precise
measurements of the CMB temperature fluctuations in this $\ell$ range
(barring problems with unanticipated systematics), and higher
$\ell-$ranges begin to be dominated by foreground sources. If there
are features in the Planck data that can not be adequately explained
by $\Lambda$CDM (such as a strong correlation between different
multipoles), we could and should devise a revised concordance
model. But we would be unable to test its predictions with future CMB
temperature measurements!

After the COBE experiment \citep{Smoot:1992td} observed hints of a low
quadrupole, it took subsequent confirming measurements by WMAP to
establish this, \citep{Spergel:2003cb,Spergel:2006hy,Komatsu:2008hk}
and to detect the planarity of the quadrupole and octopole and their
alignment with each other, perpendicular to the ecliptic, with an axis
toward the CMB dipole
\citep{deOliveiraCosta:2003pu,Schwarz:2004gk,Land:2005ad}, where cosmic
variance already is the limiting factor. Thus possible new models
explaining the low $\ell$ multipole alignments cannot be tested on
their predictions for future measurements of these multipoles.
Instead, one has to look for different predictions from the new
models, e.g. by looking for circles in the sky as a signature of a
topologically non-trivial universe \citep{Cornish:2003db}.  If only
parts of the WMAP data had been released, tantalizing enough to induce
people to look for new models, there would have been room to test the
predictions of these models for the low $\ell$s.

In the (perhaps not so distant) future, a similar situation will arise
with other cosmological experiments. Large scale structure
observations by way of galaxy counts will eventually measure the
positions and redshifts of all galaxies in the our Hubble patch with
high precision (neglecting uncertainties due to non-linearities). The
distribution of hydrogen will be mapped with observations or the
Ly-$\alpha$ forest. Eventually all observations on cosmological scales
will reach the cosmic variance limit, as we only have this one
universe from which to sample.

In light of this, it seems imperative to reflect on ways to extract an
optimal amount of information from complete finite data sets. They
should be not only be used to better constrain parameters of the
concordance model, but to discover and test new models. We need to
devise schemes for incremental data release as cosmological analogues
of blind analysis, a procedure often used in particle physics, where
the need to avoid the (possibly unconscious) influence of the
statistical methodology adopted on the significance of the results is
a well recognized problem, see e.g.~\cite{Louis}. For example, one
wants to avoid (unwillingly) biassing the significance of a signal
when designing the ``cuts'' on the number of observed events. Several
strategies have been devised to this end. For example, a random number
can be added to the data, and subtracted only after all corrections
and other data manipulations have been performed; or just a fraction
of the data is employed to define the statistical procedure, while the
remainder of the data are only revealed in a subsequent phase. After
that point no further adjustments of the methodology are allowed. The
split of data in subsets can either happen in time (an obvious
solution for many particle physics experiments) or in data space. In
the latter case, a ``signal box'' of data is left closed until
potential anomalies in the first chunk of observations have been
identified and statistical tests for their confirmation designed, at
which point the box is opened and the analysis unblinded.  An example
of such a procedure is the miniBooNE neutrino oscillation
experiment~\citep{Bazarko:2000id}. Another method is sometimes adopted
by precision measurements where the analysis team is allowed to see
the full data sets, but with arbitrary units. The resulting parameter
constraints are rescaled to the actual units only at the very end of
the analysis.

All of those strategies are designed with the common aim of keeping a
part of the information hidden from the first stage of the analysis,
so as to be able to exploit the full statistical power of the hidden
data upon unblinding. We now turn to the discussion of possible ways
of applying this idea in the cosmological context.

\section{Strategies for the release of partial data}

There is always a random element involved in choosing a good way to
split data, where the definition of ``good'' often depends on the
unknown anomalies one is hoping to be able to test. Suppose we throw a
single coin $2N$ times after which it is lost. The first $N$ throws
include an equal number of heads and tails, while the last $N$ tosses
are all tails. Splitting this data set in these two chunks, the first
set points towards the model of a fair coin. The second set (all
tails) raises serious doubt about this model. But we have no way of
verifying the predictions of a new model (e.g. the coin was exchanged
for an all-tails coin) as the coin was lost. Had we split the data
into four equal chunks, then after examining the third chunk we would
likely have proposed a new model of an unfair all-tails coin.  The
predictions of this new model would have been tested (and confirmed)
by the fourth chunk of data.

Two opposing forces are at play when considering ways to release
partial data. On the one hand, releasing individual data points will
lead to many statistical flukes that can be mistaken for features in
the data. On the other hand, releasing all data at once will only
allow to determine the parameters of the existing models and not to
check predictions of potential new models. It seems hard to find an
optimal number of chunks, even more so as it is not even clear how
data should be split.

The most natural way to release partial data is often by time
ordering, such as is employed by many experiments, for example WMAP.
A natural cut-off between data sets is the point in time when (if) the
doubt \citep{Starkman:2008py} on a concordance or reference model
reaches a critical threshold, after which an alternative model should
be devised. Using only data that was not used to compute the doubt on
the original model, compute the doubt on the new model. Iterate this
process until all data has been taken or funding runs out. This method
does not detect all features as the likelihood function typically does
not incorporate all predictions of the original model. For example,
the riddles of why the two point correlation function of the
temperature fluctuations vanishes at separation angles larger than
$60^\circ$ \citep{Copi:2005ff} and of the alignments of the quadrupole
and octopole \citep{deOliveiraCosta:2003pu,Schwarz:2004gk,Land:2005ad}
would escape detection as the likelihood function is insensitive to
these features.

Summary statistics for CMB measurements often are presented in the
form of (binned) $C_\ell$'s building on isotropy and Gaussianity of
the $a_{lm}$'s. Other quantities, such as $C(\theta)$, would work as
well. A possible course of action would be to exclusively release
binned $C_{\ell}$'s in the first data release. Then a search for
deviations from the concordance model -- new features -- could be
conducted. If any unexpected features are noted in the data, new
models would be devised and their predictions for the unbinned
$C_{\ell}$'s could be compared against the second, unbinned data
release. One might envision performing a finer graining of the binning
process, going from e.g. $\Delta \ell=10$ bins in the first year to
$\Delta \ell=5$ bins in the second year to $\Delta \ell = 1$ bins in
the third year, or in terms of the two-point function $C(\theta)$
using averaged values over $\Delta\theta=10^\circ, 1^\circ, 0.1^\circ,
\dots$ for each release cycle. A possible complication is the fact
that the successive data releases include the previous data and hence
are correlated.

However, there is a way to split data guaranteeing uncorrelated data
chunks: principal component analysis (PCA)\citep{Huterer:2002hy}.  Each
principal component, i.e. eigenvector and eigenvalue of the data's
covariance matrix, is released separately, giving as many attempts at
finding new models as there are well-constrained PCAs. Their order
seems to be a matter of taste. Releasing the best-constrained
component first would make it easiest to detect any features, then
using the less-well constrained modes to verify any new model. Not
producing any hints at a new model, this procedure -- as any splitting
of data -- would not have any negative impact on parameter estimation
(as Bayesian updating of posterior pdfs does not care about the order
of the information being added).

Independent of how the data is split, sizing the individual chunks
also seems to be rather an art. They should neither be too small,
i.e. not so noisy as to induce spurious features, nor too large, or
new models will not be testable. It may prove beneficial to release
data chunks with the same information content, as measured e.g.~by the
mean square error or an information-theory based measure such as the
Kullback--Leibler divergence.

\section{Conclusions} Cosmologists are in a paradoxical situation. They
strive to acquire data of the highest possible quality to constrain
parameters of their models as quickly as possible. But they should be
open to new features in the data that are not predicted by current
models, and hence to the possibility of having to devise new models
and test their predictions.  We have argued that for the latter step,
availability of fresh data is crucial, which for cosmic variance
limited data sets is simply not possible.  We therefore propose that
such {\it ultimate} data sets be treated as the precious resources
they are and released slowly and carefully.

We have discussed various strategies for parsing such data sets. It
remains an art to find the optimal way to split data and release it,
involving inevitably a certain degree of luck to detect unexpected
features.  It seems to us from this first overview that the most
promising way of ``dividing the plunder'' is to employ a PCA
decomposition of the data and release data parts of equal information
content.  This is a compromise between being able to find new features
and having enough data left to reliably test possible new models.
However, the best strategy is likely to depend heavily on the
particular data set, and on the taste of the individual investigators.
Wishing to avoid that basic human failing of over-confidence we
acknowledge that there is a reasonable chance that we have overlooked
the optimal strategy.

We urge our observational colleagues to be frugal with their data.
Slicing the data and doling it out slowly is in all of our long term
best interests.

\bigskip 
{\em Acknowledgements:} We would like to thank Andrew Jaffe, Louis
Lyons, Irit Maor and John Ruhl for stimulating discussions. G.D.S. and
P.M.V. are supported by a grant from the DOE to the
particle-astrophysics group at CWRU. P.M.V. is supported by the office
of the Dean of the College of Arts and Sciences at CWRU. R.T. would
like to thank CWRU for hospitality. We thank E. Scrooge for
motivation.  \bibliographystyle{mn2e} \bibliography{frugality}
\end{document}